\documentclass[11pt]{article}

\usepackage[]{acl}

\usepackage{times}
\usepackage{latexsym}
\usepackage{soul}

\usepackage[T1]{fontenc}
\usepackage[utf8]{inputenc}

\usepackage{microtype}

\usepackage{inconsolata}

\usepackage{graphicx}

\title{\sys: Breaking Topology Confidentiality in LLM Multi-Agent Systems with Stealthy Context-Based Inference}
\author{
Zixun Xiong\textsuperscript{1},
Gaoyi Wu\textsuperscript{1},
Lingfeng Yao\textsuperscript{2},
Miao Pan\textsuperscript{2}, \\
\textbf{Xiaojiang Du}\textsuperscript{1},
\textbf{Hao Wang}\textsuperscript{1} \\
\textsuperscript{1}Department of Electrical and Computer Engineering, Stevens Institute of Technology \\
\textsuperscript{2}Department of Electrical and Computer Engineering, University of Houston \\
\texttt{\{zxiong9, gwu13, xdu16, hwang9\}@stevens.edu} \\
\texttt{\{lyao12, mpan2\}@uh.edu}
}

\usepackage{hyperref}
\usepackage{xspace}
\acrodef{LLMMAS}[LLM-MAS]{LLM-based multi-agent system}
\newcommand{\LLMMAS}{\ac{LLMMAS}\xspace}

\acrodef{DDPM}[DDPM]{denoising diffusion probabilistic model}
\newcommand{\DDPM}{\ac{DDPM}\xspace}

\acrodef{GCG}[GCG]{greedy coordinate gradient}
\newcommand{\GCG}{\ac{GCG}\xspace}

\acrodef{MRR}[MRR]{mean reciprocal rank}
\newcommand{\MRR}{\ac{MRR}\xspace}

\acrodef{SOTA}[SOTA]{state-of-the-art}
\newcommand{\SOTA}{\ac{SOTA}\xspace}

\acrodef{IP}[IP]{intellectual property}
\newcommand{\IP}{\ac{IP}\xspace}

\usepackage{acronym}
\usepackage{algorithm}
\usepackage{algpseudocode}
\usepackage{amsmath}
\usepackage{amssymb}

\usepackage{multirow} 
\usepackage{booktabs}  % 提供 \toprule, \midrule, \bottomrule 等
\usepackage{makecell}
\usepackage{tcolorbox}
\newcommand{\sys}{\textit{WebWeaver}\xspace}
\renewcommand{\Return}{\State \textbf{return}}

\definecolor{ForestGreen}{RGB}{34,139,34}
\definecolor{RoyalBlue}{rgb}{0.25, 0.41, 0.88}

\newtcolorbox{promptbox}{
    colback=gray!5!white,   % 背景色：极浅的灰色 (5% gray)
    colframe=black,         % 边框颜色：黑色
    boxrule=0.5pt,          % 边框粗细：细线
    arc=2mm,                % 圆角弧度：2mm (对应截图的微圆角)
    boxsep=1mm,             % 内容与边框的间距
    left=2mm, right=2mm, top=1mm, bottom=1mm, % 内部边距
    fontupper=\small\rmfamily % 字体：稍微缩小，使用罗马字体(Serif)以匹配截图
}
\begin{document}
\maketitle
\begin{abstract}
    Communication topology is a critical factor in the utility and safety of LLM-based multi-agent systems (LLM-MAS), making it a high-value intellectual property (IP) whose confidentiality remains insufficiently studied.
    Existing topology inference attempts rely on impractical assumptions, including control over the administrative agent and direct identity queries via jailbreaks, which are easily defeated by basic keyword-based defenses. As a result, prior analyses fail to capture the real-world threat of such attacks.
    To bridge this realism gap, we propose \textit{WebWeaver}, an attack framework that infers the complete LLM-MAS topology by compromising only a single arbitrary agent instead of the administrative agent. 
    Unlike prior approaches, WebWeaver relies solely on agent contexts rather than agent IDs, enabling significantly stealthier inference.
    WebWeaver further introduces a new covert jailbreak-based mechanism and a novel fully jailbreak-free diffusion design to handle cases where jailbreaks fail. 
    Additionally, we address a key challenge in diffusion-based inference by proposing a masking strategy that preserves known topology during diffusion, with theoretical guarantees of correctness.
    Extensive experiments show that WebWeaver substantially outperforms state-of-the-art (SOTA) baselines, achieving about 60\% higher inference accuracy under active defenses with negligible overhead.
\end{abstract}

\section{Introduction}
\begin{figure}
    \centering
    \includegraphics[width=\linewidth]{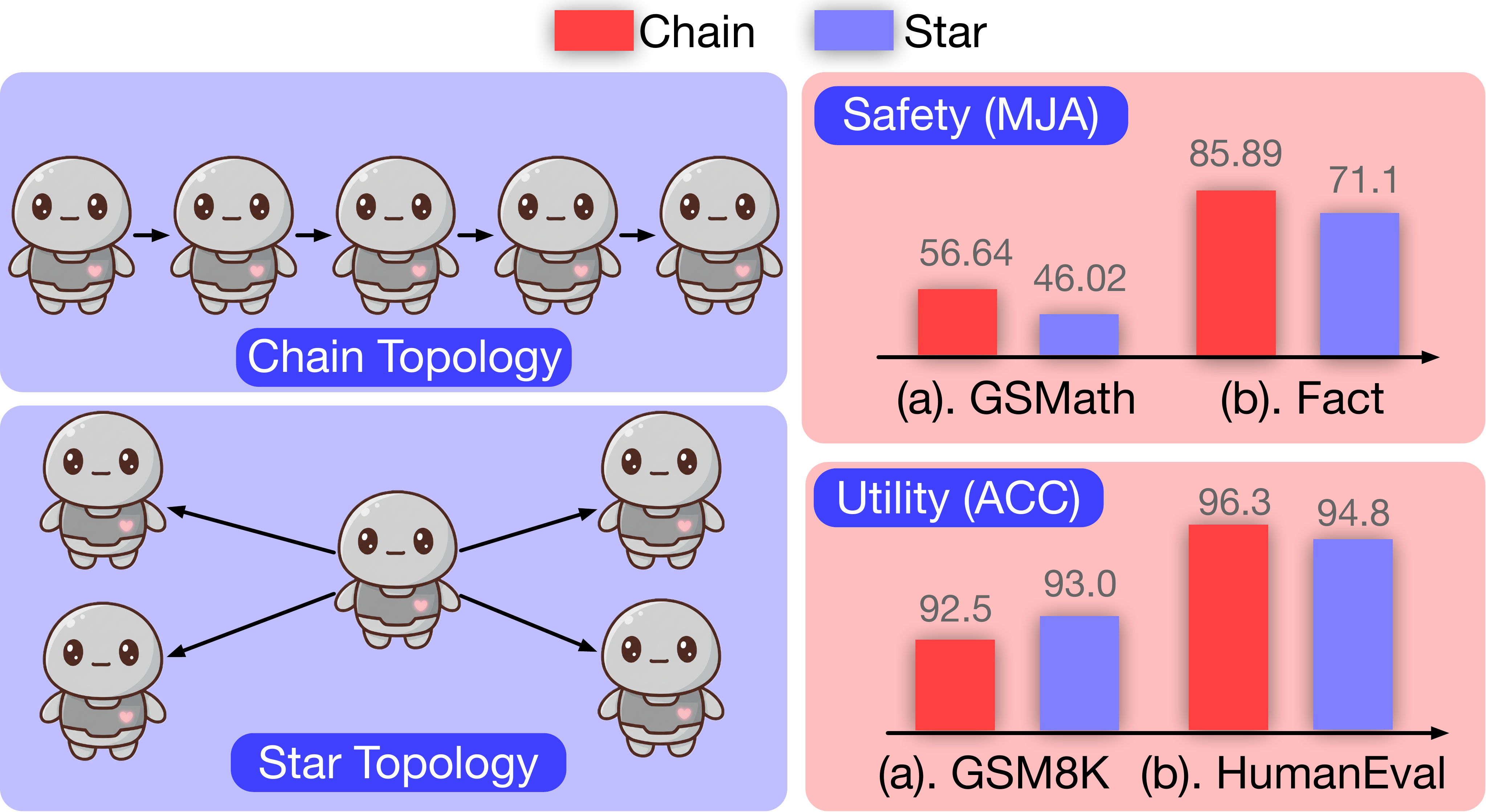}
    \caption{Studies~\cite{yu2024netsafe,yang2025topological} show that communication topology in \LLMMAS significantly affects security and utility performance, with no universally optimal topology, making it a critical form of intellectual property.}
    \label{fig: intro_topo}
\end{figure}

% Recently, 
\Acp{LLMMAS} have demonstrated expert-level proficiency across scientific~\cite{lu2024ai, schmidgall2025agent, gottweis2025towards} and industrial domains~\cite{hong2023metagpt, kim2025atlantis, guan2025rstar}, proving highly effective for complex reasoning tasks.
Beyond individual agent capabilities, empirical evidence highlights the \textit{communication topology} as a decisive performance factor~\cite{yu2024netsafe, zhang2025stop}; variations in topology design yield significant disparities in system utility~\cite{zhang2025stop, yang2025topological, cemri2025multi} and safety~\cite{zhang2025stop} even with identical constituent agents, positioning optimized topologies as high-value \IP as Figure~\ref{fig: intro_topo} shows. 
Moreover, prior work indicates that adversaries possessing knowledge of this structure can execute significantly more sophisticated attacks~\cite{shahroz2025agents, wang2025g} compared to structure-agnostic strategies. Despite these risks, the confidentiality of multi-agent topologies remains underexplored, necessitating a systematic investigation into topology inference attacks to address this critical security gap.

The limited prior work on topology inference suffers from impractical assumptions, leaving the real-world threat of such attacks insufficiently understood.
Specifically, previous work assumes that the attacker controls the administrative agent responsible for initiating the system~\cite{wang2025ip}. However, this assumption is unrealistic in collaborative \LLMMAS settings where different entities operate different agents~\cite{lu2024ai, schmidgall2025agent, gottweis2025towards}. In practice, it is more reasonable to assume that the attacker can compromise a single arbitrary agent rather than the administrative agent.
Moreover, existing approaches infer topology by directly querying agent identities via jailbreaks, which can be easily defeated by basic keyword-based defenses.

To address these limitations, we introduce \sys, a new attack framework designed to infer \LLMMAS topologies. Distinct from prior works that necessitate control over the system administrator, \sys operates by compromising a single arbitrary agent. 
\sys further introduces a covert recursive jailbreak-based mechanism, with a fully jailbreak-free masked diffusion model as a fallback topology completion module. 
Since they do not need agent IDs and are not based on certain keywords, \sys is robust to keyword-based defenses.
Additionally, we address a key challenge in the diffusion module by proposing a masking strategy that preserves the known topology during diffusion, with our theoretical guarantees of correctness.
Our contributions are as follows: 
\begin{itemize}
    \item We identify a key realism gap in prior topology inference work and propose \sys, the first framework to recover complete \LLMMAS topologies by compromising a single arbitrary agent---without requiring administrative privileges or relying on impractical jailbreaks to directly extract neighboring agents’ IDs. 

    \item We also construct a dialogue dataset with explicitly annotated topology, agent prompts, and sender-receiver labels, which facilitates both our study and future research on the security of topology in \LLMMAS.
    
    \item We propose a more stealthy attack that infers LLM-MAS topology from agent contexts rather than identifiers, combining a covert recursive jailbreak mechanism with a jailbreak-free diffusion module, and addressing the new challenge of ensuring correctness via a masking strategy that preserves known topology
    
    \item Extensive evaluations across diverse datasets demonstrate that \sys outperforms \SOTA baselines by approximately 60\% in inference accuracy under active defenses.
    Furthermore, both methods maintain negligible overhead, with the jailbreak-free version incurring zero additional computational cost on the target system.
\end{itemize}

% \begin{itemize}
%     \item TODO: two methods (jailbreak-based and jailbreak-free version)
%     \item 
% \end{itemize}
\section{Related Works}
\paragraph{\LLMMAS Topology.}
% \Acp{LLMMAS}, collaborative frameworks where multiple autonomous agents powered by large language models interact through a defined communication topology, have achieved expert-level performance across scientific and industrial domains. 
% %
% In the realm of scientific discovery, \citet{lu2024ai} introduce a framework for fully automated discovery, while \citet{schmidgall2025agent} and \citet{gottweis2025towards} develop autonomous research assistants and co-scientists to optimize laboratory workflows and collaborative experimentation. 
% %
% For industrial tasks, \citet{hong2023metagpt} incorporate standardized operating procedures into multi-agent coordination, \citet{kim2025atlantis} utilize MAS for autonomous cyber-threat localization and triage, and \citet{guan2025rstar} enhance complex mathematical reasoning via self-evolved deep thinking. Beyond task-specific capabilities, empirical research highlights the communication topology as a decisive factor in system utility and safety~\cite{yu2024netsafe, zhang2024psysafe, yang2025topological}, with \citet{cemri2025multi} identifying specific failure modes arising from structural misalignments. 
% %
% While optimized topologies are recognized as high-value proprietary assets, adversaries possessing structural knowledge can launch significantly more sophisticated attacks~\cite{shahroz2025agents} compared to structure-agnostic strategies. 
% %
% Despite these risks, the confidentiality of multi-agent topologies remains critically underexplored. \hnote{this paragraph could be condensed if lack of space}
\Acp{LLMMAS} are collaborative systems in which autonomous agents powered by large language models interact through a communication topology, and they have shown strong performance across scientific and industrial domains~\cite{lu2024ai, schmidgall2025agent, gottweis2025towards, hong2023metagpt, kim2025atlantis, guan2025rstar}.
Recent studies show that communication topology is critical to system utility and safety, with structural issues leading to concrete failures~\cite{yu2024netsafe, zhang2024psysafe, yang2025topological, cemri2025multi}. As optimized topologies are valuable assets, adversaries with structural knowledge can launch stronger attacks than topology-agnostic ones~\cite{shahroz2025agents}.
However, the confidentiality of multi-agent communication topologies remains largely unexplored.

\begin{figure*}[t]
    \centering
    \includegraphics[width=0.9\linewidth]{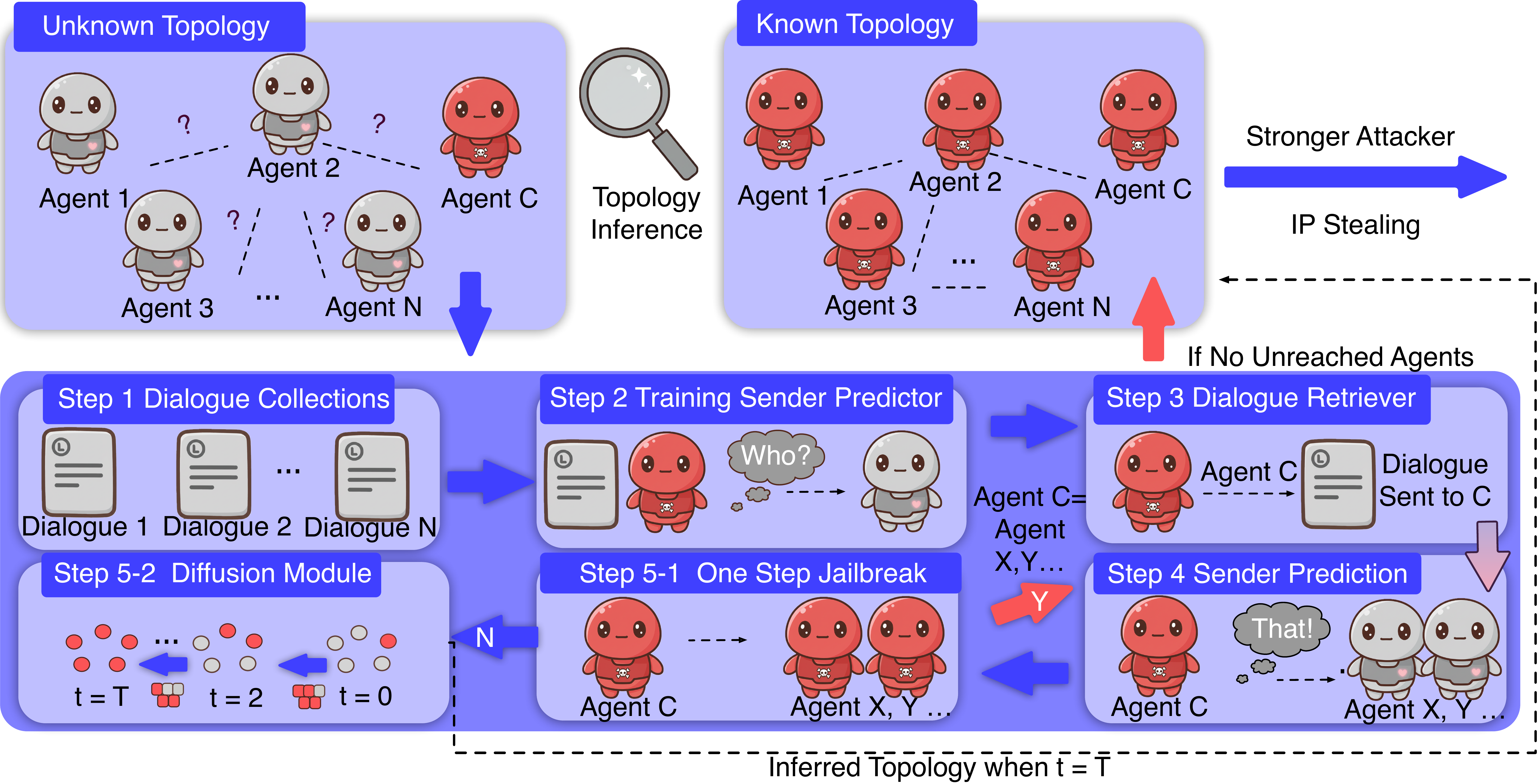}
    \caption{The pipeline of \sys: \textbf{Step-1:} Inter-agent dialogues are collected offline under known topologies and reused for both training and inference. 
\textbf{Step-2:} The collected dialogues are used to train a sender predictor that infers the sender identity from a received dialogue. 
\textbf{Step-3:} The attacker compromises a selected agent $C$ and retrieves all dialogues received by this agent during interaction. 
\textbf{Step-4:} The trained sender predictor is applied to the retrieved dialogues to infer the set of agents directly connected to the selected agent. 
\textbf{Step-5:} If the initial or optimized jailbreak succeeds (denoted as Y), the attacker uses it to induce connected agents to request additional context and iteratively repeats dialogue retrieval and sender prediction to expand the inferred graph until no new agents are uncovered; if the optimized jailbreak still fails (denoted as N), the attacker performs topology completion using a masked diffusion model (i.e., DDPM) trained on the collected dialogues and conditioned on the partially known topology to infer the complete interaction structure.} 
    \label{fig:pipeline}
\end{figure*}

\paragraph{Vulnerabilities in \LLMMAS .} 
% The security of \LLMMAS has been scrutinized across various attack vectors, ranging from availability to behavioral integrity. 
% %
% For instance, CORBA~\cite{zhou2025corba} explores the vulnerability of these systems to contagious recursive blocking attacks that can cause widespread denial of service.
% %
% In the realm of behavioral safety, PsySafe~\cite{zhang2024psysafe} provides a comprehensive framework for psychological-based attacks, defenses, and evaluations.
% %
% \hnote{can we categorize attacks and defenses, and \textbf{show our position} clearly?}
% To counter such threats, recent works have proposed specialized defenses: BlockAgents~\cite{chen2024blockagents} utilizes blockchain technology to ensure Byzantine-robust coordination, while AgentSafe introduces hierarchical data management to safeguard inter-agent communication.
% %
% Despite these advancements, topology inference attacks represent a distinct and underexplored threat aimed at compromising \LLMMAS topology confidentiality.
% %
% Existing research~\cite{wang2025ip} in this niche is largely limited to IP leakage attacks, which suffer from a significant realism gap. 
% %
% Specifically, prior work~\cite{wang2025g} typically relies on directly querying neighbor agent IDs via jailbreaks, a strategy easily nullified by defenses that audit messages for agent indices, and often assumes the attacker maintains unrealistic control over the agent that initiates the \LLMMAS, which is rarely the case in collaborative multi-institute environments~\cite{lu2024ai, schmidgall2025agent, gottweis2025towards}.
Prior work on \LLMMAS security can be grouped into three categories: system availability, behavioral safety, and topology confidentiality.
The first two study denial-of-service attacks, psychological manipulation, and defenses that protect agent behavior and communication~\cite{zhou2025corba, zhang2024psysafe, chen2024blockagents}.
The third focuses on inferring the communication topology but remains limited and mainly treats topology leakage as an intellectual property issue~\cite{wang2025ip}.
Existing approaches in this category rely on unrealistic assumptions, such as direct identity queries via jailbreaks and control over the initiating agent, which rarely hold in collaborative settings~\cite{lu2024ai, schmidgall2025agent, gottweis2025towards}.
Our work falls into the third category and studies topology inference under a realistic threat model based on a single agent and context-based inference.
\section{Proposed Methods}

 % (TODO: add reference to all related sections)

\subsection{Threat Models}
\label{sec: threat_model}
We assume the target \LLMMAS consists of $N$ agents that may be arranged in different topological structures, among which $k$ agents are compromised ($1\le k \le \frac{N}{2}$).
We also assume that the system contains no isolated agents, as agents without connections cannot contribute to the \LLMMAS. Thus, we set $k=1$ in all experiments.
\paragraph{Attacker's Goal.}
The attacker aims to infer the communication topology of a target \LLMMAS, which is a high-value intellectual property of its owner. Knowledge of the topology also enables the attacker to better understand agent coordination and to launch stronger downstream attacks~\cite{shahroz2025agents, wang2025g}. This objective naturally arises in collaborative settings such as inter-university research~\cite{lu2024ai, schmidgall2025agent, gottweis2025towards}, where institutions deploy their own agents and interact through shared tasks. In such scenarios, an institution may infer the topology of the target \LLMMAS from observed interactions. As a result, it can steal the intellectual property of the \LLMMAS owner and deploy more advanced attacks to access the private data of other agents.

\paragraph{Attacker's Ability.}
We consider a realistic attacker whose capabilities are consistent with the example.
First, the attacker controls a single arbitrary agent in the target \LLMMAS and can proactively request context from adjacent agents. This behavior is reasonable in practice, as agents often operate on different devices and may request context retransmission due to asynchronous execution or temporary disconnections.
Second, the attacker knows which agents participate in the target \LLMMAS and can collect inter-agent dialogues using a separate \LLMMAS under its own control. 
For example, in collaborations between universities, each institution may host its own agent specialized in a particular domain. A university focusing on \texttt{Domain A} may operate its own \LLMMAS to collaborate with other universities and record dialogue exchanges during joint research activities~\cite{lu2024ai, schmidgall2025agent, gottweis2025towards}. 
These collected dialogues can then be leveraged to infer the communication topology of another \LLMMAS deployed by a different university specializing in \texttt{Domain B}, thereby achieving the attacker’s goal.

\paragraph{Challenges and Contributions.}
1). Topology inference in \LLMMAS often relies on explicitly querying agent IDs, which is easily defeated by keyword-based defenses; we instead propose a training-based framework that infers topology purely from contextual signals inherent in multi-agent collaboration, eliminating the need for identifier queries.
2). Existing jailbreak strategies are largely static and thus detectable in multi-agent settings; we introduce an adaptive jailbreak that dynamically adjusts its prompts, achieving substantially higher stealthiness under adaptive defenses.
3). When jailbreaks fail, topology inference reduces to a zero-shot, NP-hard graph completion problem~\cite{fang2025contrastive}; we model this setting as diffusion-based graph reconstruction using dynamic dialogues, while addressing the new challenges of structural consistency to preserve known topology during inference.

\subsection{Jailbreak-based Module}
\paragraph{Data Collection \& Sender Predictor Training.}
The extraction process begins by passively collecting interaction logs from the compromised \LLMMAS environment as discussed in Figure~\ref{fig:pipeline}. 
%
% \hnote{can we claim this dataset as our contribution?}
%
We compile a dataset of dialogue history $\mathcal{H}$, where each entry consists of message content and the associated sender identity. For example, a log sequence might appear as: \textit{\{sender: Agent A, receiver: Agent B, dialogue: ``XXX''\}}.
Using this labeled dataset~\footnote{We will release the dataset recently.}, we train a sender predictor, $S_{\theta}$, to learn the distinct linguistic fingerprints and role-specific syntax of different agents.
The predictor is optimized to map a message content snippet $m$ to a specific agent identity $s$, maximizing the probability:
$P_{\theta}(\text{sender}=s | m).$
This allows us to de-anonymize the source of a message based purely on its semantic content.

\paragraph{Dialogue Retrieval \& Sender Prediction.}
Once trained, we deploy the $S_{\theta}$ in conjunction with a compromised agent, denoted as $A_{C}$. 
This agent acts as a listener within the graph. As $A_{C}$ receives incoming messages from its immediate neighbors during standard system operation, it feeds these dialogue fragments into $S_{\theta}$.
For every received message $m_{i}$, the predictor outputs the most probable sender identity:
$\hat{s} = \arg\max_s S_{\theta}(m_{i}).$ 
By aggregating these predictions over time, $A_{C}$ effectively infers its local adjacency matrix $\mathbf{A}_{obs}$, identifying all agents $A_i$ such that a direct edge $(A_i, A_{C})$ exists, solely by analyzing the ``voice'' of the incoming traffic.

\paragraph{Recursive Jailbreak.}
To expand this discovery from the local neighborhood to the whole graph, we implement a recursive jailbreak mechanism. We inject a ``propagation prompt'' into $A_{C}$, instructing it to command its neighbors to:
1). forward their own conversation histories (without the sender index) back to $A_{C}$, 
and 2). repeat this command to \textit{their} neighbors. This creates a cascade of context leakage, allowing us to recursively infer the global topology with the help of the sender predictor. 
However, if neighbors are equipped with safety filters that reject this instruction, we escalate to an optimization-based attack.
We employ a \GCG approach using a local proxy LLM. Since we lack white-box access to the victim agents, we calculate gradients on the local model to optimize an adversarial suffix that maximizes the likelihood of compliance. 
To formalize this, we define the optimization objective as minimizing the negative log-likelihood of the target response:
\begin{equation}
    \min_{\boldsymbol{\delta}} \mathcal{L}(\boldsymbol{\delta}) = - \sum_{t=1}^{|P|} \log p(y_t \mid \mathbf{x}_{prop} \oplus \boldsymbol{\delta}),
    \label{eq: suffix_optim}
\end{equation}
where $y_t$ is a sentence that starts with ``Sure'', $x$ is the original ``propagation prompt'', $\delta$ is the suffix we want to optimize, and $\oplus$ is the concatenation operation.
We solve this using the \GCG method by computing gradients with respect to token embeddings of $\delta$:
$    \nabla_{{\delta_i}} \mathcal{L} \quad \text{for } i = 1, \dots, |\boldsymbol{\delta}|.$
Finally, we iteratively update the suffix by selecting the best candidate from the top-$k$ substitutions
This optimized suffix is then transferred to the target system. By iteratively refining the prompt until it bypasses the defense, we force the neighbors to execute the propagation command, eventually retrieving the global context and revealing the entire network structure $\hat{G}$.
The complete procedure is summarized in the Appendix.
\subsection{Jailbreak-free Diffusion Module}
We also propose a diffusion-based module to infer the global topology of the target \LLMMAS when jailbreak is infeasible.
We draw inspiration from diffusion models, which are fundamentally designed for denoising.
Moreover, our task can be viewed as a graph completion problem, where topology inference naturally corresponds to denoising a partially observed interaction graph. 
We also address a key challenge in applying DDPMs, namely that standard diffusion processes cannot preserve known topology; to resolve this, we introduce a masking module, as described later.\footnote{The proof of why our masking can be applied without violating the properties of DDPM is provided in the Appendix.}
Since we collect a set of interaction topologies and associated dialogues that can be used for training, we model topology inference as a diffusion process, enabling effective inference (stealing) of the underlying interaction topology.
Thus, only Steps 1 and 2 in Figure~\ref{fig:pipeline} are required when combined with the diffusion model, eliminating the need for jailbreak.

\paragraph{Problem Formulation.}
We represent the \LLMMAS topology as an undirected graph $\mathcal{G} = (\mathcal{V}, \mathcal{E})$, characterized by a symmetric adjacency matrix $\mathbf{A} \in \{0, 1\}^{N \times N}$ describing both vertices (i.e., agents) $V$ and edges $E$, where $N$ denotes the number of agents. 
We also consider a scenario of partial observability where connections for a subset of agents are known~\footnote{This is done by Step~2 and Step~3}: 
Let $\mathbf{M} \in \{0, 1\}^{N \times N}$ be a binary mask matrix, where $M_{ij} = 1$ indicates that the relationship between agent $i$ and agent $j$ is observed, and $M_{ij} = 0$ indicates an unknown status. Given the observed partial topology $\mathbf{A}_{obs} = \mathbf{A}_{gt} \odot \mathbf{M}$, our goal is to infer the ground-truth topology $\mathbf{A}_{gt}$ by inferring the missing entries in $(\mathbf{1} - \mathbf{M}) \odot \mathbf{A}_{gt}$, where $\odot$ denotes the Hadamard product.

\paragraph{Diffusion-based Topology Inference Generative Model.}
The aforementioned problem can be addressed using a diffusion model: the observed subgraph is treated as a noisy input $\mathbf{A}_{obs}$, while the ground-truth graph corresponds to the clean signal $\mathbf{A}_{gt}$. Since diffusion models are designed to denoise corrupted inputs, they can be leveraged to recover $\mathbf{A}_{gt}$ too.
All we need to do in this part is to model the forward and reverse processes of the diffusion model.
Specifically, we employ a \DDPM to learn the distribution of target topology, where $\mathbf{x}_0 = \mathbf{A}_{gt}$. The forward process is a Markov chain that gradually adds Gaussian noise to the data according to a variance schedule $\beta_1, \dots, \beta_T$:
\begin{equation}
    q(\mathbf{x}_t | \mathbf{x}_{t-1}) = \mathcal{N}(\mathbf{x}_t; \sqrt{1-\beta_t}\mathbf{x}_{t-1}, \beta_t \mathbf{I}).
\end{equation}
The reverse process learns to denoising the graph structure by predicting the noise $\boldsymbol{\epsilon}_\theta(\mathbf{x}_t, t)$ to approximate the posterior $q(\mathbf{x}_{t-1}|\mathbf{x}_t)$. The sampling step is defined as:
\begin{equation}
    \mathbf{x}_{t-1} = \frac{1}{\sqrt{\alpha_t}} \left( \mathbf{x}_t - \frac{1-\alpha_t}{\sqrt{1-\bar{\alpha}_t}} \boldsymbol{\epsilon}_\theta(\mathbf{x}_t, t) \right) + \sigma_t \mathbf{z},
    \label{eq:reverse_step}
\end{equation}
where $\alpha_t = 1 - \beta_t$, $\bar{\alpha}_t = \prod_{i=1}^t \alpha_i$, and $\mathbf{z} \sim \mathcal{N}(\mathbf{0}, \mathbf{I})$.

\paragraph{Topology Inference via Subgraph Guidance.}
Before applying the above \DDPM for graph completion, a key challenge arises: directly using \DDPM would inevitably corrupt the known subgraph (the observed topology).
To address this issue, we introduce a \textit{Masked Topology Inpainting} strategy during the reverse sampling phase. Standard unconditional sampling may generate valid topologies that are inconsistent with the observed nodes. To enforce consistency, we fuse the model's prediction with the observed ground truth at each denoising step.

Specifically, at each reverse step $t-1$, we obtain the model's prediction $\mathbf{x}_{t-1}^{pred}$ using Equation~\eqref{eq:reverse_step}. Simultaneously, we compute the noisy state of the observed data, denoted as $\mathbf{x}_{t-1}^{obs}$, by sampling from the forward process marginal $q(\mathbf{x}_{t-1}|\mathbf{x}_0)$ using the known $\mathbf{A}_{obs}$\footnote{Directly injecting clean ground-truth data would cause a distribution shift, as the model expects noisy inputs at intermediate steps. Equation~(3) ensures the known subgraph carries the correct noise level corresponding to timestep $t-1$, maintaining consistency with the diffusion trajectory.}:
$$
    \mathbf{x}_{t-1}^{obs} \sim \mathcal{N}(\sqrt{\bar{\alpha}_{t-1}} \mathbf{A}_{obs}, (1 - \bar{\alpha}_{t-1})\mathbf{I}).
$$
Critically, we inject the known information into the sampling trajectory using the mask $\mathbf{M}$:
$$
    \mathbf{x}_{t-1} = (\mathbf{1} - \mathbf{M}) \odot \mathbf{x}_{t-1}^{pred} + \mathbf{M} \odot \mathbf{x}_{t-1}^{obs}.
$$
This operation ensures that the known subgraph region follows the correct noise trajectory grounded in $\mathbf{A}_{obs}$, while the diffusion model fills in the masked regions $(\mathbf{1}-\mathbf{M})$ to form a coherent global structure. 
After $T$ epochs, we can derive the predicted topology.
The complete procedure is summarized in Algorithm~\ref{alg:inpainting}.

\section{Evaluation}
\paragraph{Datasets \& Models}
We evaluate our framework on four diverse datasets. CSQA~\cite{talmor2019commonsenseqa} contains 12,247 multiple-choice questions requiring commonsense reasoning over ConceptNet, while GSM8k~\cite{cobbe2021training} includes 8,500 grade-school math problems that test multi-step logical deduction. Fact~\cite{yu2024netsafe} is a synthetic dataset of indisputable factual statements generated via GPT-4 to evaluate information fidelity, and Bias~\cite{yu2024netsafe} consists of stereotype-based statements designed to assess safety and alignment within agent topologies.
We consider four representative models: Llama 3.1 8B~\cite{dubey2024llama}, Qwen 2.5 7B~\cite{qwen2024qwen25}, Mistral NeMo 12B~\cite{mistral2024nemo}, and Gemma 2 9B~\cite{team2024gemma2}. For each experiment, a single model is uniformly assigned to all agents, and final results are averaged over multiple \LLMMAS configurations spanning all four models.

\paragraph{Agents and \LLMMAS.} For each agent, there's a different prompt for their roles\footnote{For details of the prompts, please refer to the appendix.}.
Considering previous work encourages different roles and model heterogeneity~\cite{, lu2024ai, schmidgall2025agent, gottweis2025towards}, our design is practical. 
\paragraph{Baselines.}To ensure that our results are representative, in addition to directly comparing with the only baseline~\cite{wang2025ip} that is directly relevant to our task, we also include the most recent optimization-based graph completion method. Specifically, we adopt CZRL~\cite{fang2025contrastive}, which infers the entire topology directly from the observed nodes and edges using a model trained on collected data (we assume the attacker can access dialogues from other agents via the jailbreak and use dialogues to learn the text embeddings of each node).
\paragraph{Metrics.} We use \MRR to evaluate the accuracy of the attacker in inferring the underlying \LLMMAS communication topology. 
For each agent, the attacker ranks all candidate agents as potential neighbors, and MRR measures how highly the true connected agents are ranked in this list. 
This metric is well-suited for topology inference since it directly reflects the quality of the predicted ranking rather than a single hard decision. 
A higher MRR indicates that the attacker consistently prioritizes true communication links, demonstrating more accurate topology recovery.
We further report Precision, Recall, and F1 Score to evaluate the accuracy of both the sender predictor and the subsequent graph completion results. 
Precision measures the fraction of predicted senders or inferred links that are correct, while Recall quantifies the coverage of true senders or communication links recovered by the attacker. 
The F1-score provides a balanced assessment by jointly considering precision and recall, which is critical in the presence of class imbalance and multiple candidate agents. 
Together, these metrics complement ranking-based evaluation by characterizing the correctness of discrete sender identification and topology inference.
\subsection{Main Results}
\paragraph{Sender Prediction.}
\begin{table}[t]
\centering
\scalebox{0.8}{
    \begin{tabular}{lccc}
    \toprule
    \textbf{Dataset} & \textbf{Precision} & \textbf{Recall} & \textbf{F1 Score} \\
    \midrule
    CSQA  & 0.8794 & 0.8773 & 0.8772 \\
    GSM8k & 0.9471 & 0.9472 & 0.9470 \\
    Fact  & 0.9606 & 0.9597 & 0.9597 \\
    Bias  & 0.9830 & 0.9830 & 0.9830 \\
    \bottomrule
    \end{tabular}
}
\caption{Sender predictor performance across different datasets.}
\label{tab: sender_predictor}
\end{table}
\begin{figure}[t]
    \centering
    \includegraphics[width=\linewidth]{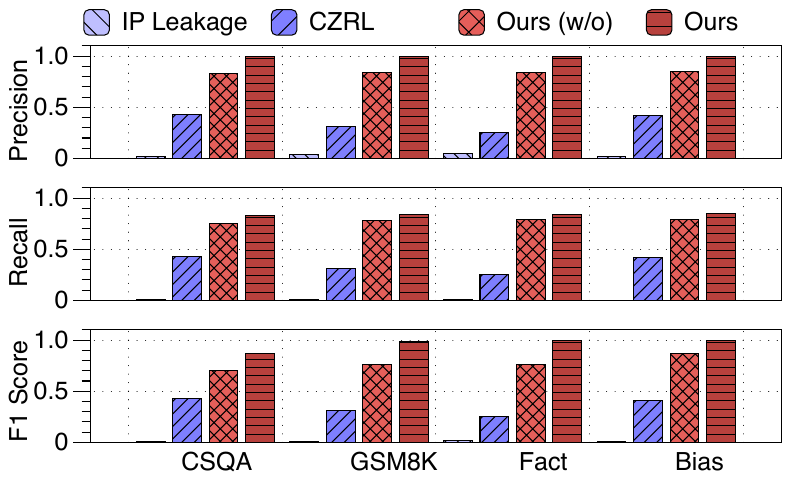}
    \caption{Robustness of different attacks against a keyword-based defense that rejects any request containing information intended to reveal adjacent agent IDs. ``Ours (w/o)'' is the jailbreak-free module of \sys, and ``Ours'' is the jailbreak-based \sys.}
    \label{fig: robustness}
\end{figure}
\begin{figure}
    \centering
    \includegraphics[width=\linewidth]{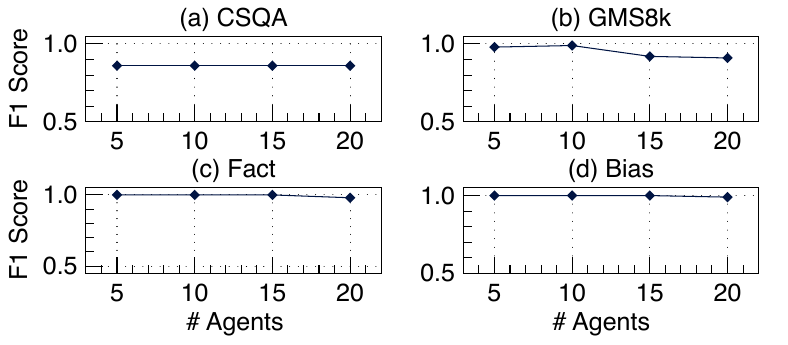}
    \caption{Scalability of the jailbreak-based module with different numbers of agents.}
    \label{fig: scale}
\end{figure}
\begin{table*}[t]
\centering
\scalebox{0.8}{
\begin{tabular}{l|cccc|cccc}
\toprule
\multirow{2}{*}{\textbf{Dataset}} 
& \multicolumn{4}{c|}{\textbf{Jailbreak-based module}} 
& \multicolumn{4}{c}{\textbf{Jailbreak-free module}} \\
& Precision & Recall & F1 Score & \MRR
& Precision & Recall & F1 Score & \MRR\\
\midrule
CSQA  
& 1.0000 & 0.7688 & 0.8666 & 0.7292 
& 0.8310 & 0.7506 &  0.7887& 0.7029\\

GSM8k 
& 1.0000 & 0.9688 & 0.9821 & 0.9208
& 0.8418 & 0.7844 &0.8120& 0.7593\\

Fact  
& 1.0000 & 1.0000 & 1.0000 & 1.0000
& 0.8432 & 0.7876 &  0.8145& 0.7618 \\

Bias  
& 1.0000 & 1.0000 & 1.0000 &1.0000
& 0.8518 & 0.7902 & 0.8198  & 0.8659\\
\bottomrule
\end{tabular}
}
\caption{Graph completion performance under jailbreak-based and jailbreak-free modules of \sys. }
\label{tab: graph_completion}
\end{table*}
To demonstrate that our sender predictor can successfully identify the sender of the given dialogue, we examine Precision, Recall, and F1 Score averaged over all dialogue traces across four different datasets.   
As shown in Table~\ref{tab: sender_predictor}, our sender predictors achieve consistently strong performance across all four datasets, demonstrating their ability to effectively distinguish among different agents.
This indicates that agents exhibit distinct behavioral or stylistic identities, which are implicitly reflected in their generated dialogues. 
As a result, the sender predictor can successfully learn discriminative patterns from dialogue content alone, enabling accurate sender identification.
Overall, the performance is robust, with all datasets achieving F1 scores above 0.85, highlighting the reliability of the proposed predictor.
We also observe that CSQA yields comparatively lower scores than the other datasets, likely due to its more diverse reasoning paths and less consistent dialogue patterns.
In contrast, datasets such as GSM8k, Fact, and Bias exhibit more structured or homogeneous dialogue characteristics, making agent-specific features easier to capture and leading to higher prediction accuracy.

\paragraph{Graph Completion.} To demonstrate the effectiveness of our graph completion capability, we evaluate the performance of topology inference under both jailbreak-based and jailbreak-free modules, as reported in Table~\ref{tab: graph_completion}.
Overall, both modules achieve strong results across all datasets, indicating that the proposed graph completion strategy is effective even without relying on jailbreak capabilities.
The jailbreak-based module consistently attains higher precision and F1 scores, in some cases reaching perfect performance, suggesting that direct context access provides more complete and reliable neighborhood information for topology inference. 
In contrast, the jailbreak-free module exhibits a moderate performance drop, particularly on CSQA, but still maintains competitive F1 scores above 0.78 across all datasets.
This demonstrates that our approach remains robust in more restrictive settings, and that sufficient structural signals can be inferred from local topology alone. 
The relatively larger gap on CSQA can be attributed to its higher reasoning diversity and noisier interaction patterns leading to the performance degradation of the sender predictor, whereas more structured datasets such as GSM8k, Fact, and Bias facilitate more accurate graph inference.
A few inference samples of the jailbreak-free module are shown in Figure~\ref{fig: topology_reconstruction}.

\paragraph{Robustness against Potential Defenses.}

Figure~\ref{fig: robustness} illustrates the robustness of different attack strategies against filtering-based defenses, where the sender agent’s index is filtered from the communication.
As observed, the IP Leakage baseline exhibits near-zero performance across all metrics; this collapse is expected, as the method relies exclusively on explicit agent indices, which are effectively neutralized by the filter. 
While CZRL maintains moderate performance, it remains sub-optimal because it fails to fully exploit the compromised agent's capability to intercept incoming dialogues and actively disseminate messages to neighbors.
In contrast, our approach demonstrates superior resilience. 
Even our jailbreak-free module (``Ours (w/o)'') outperforms the \SOTA by inferring topology solely through dialogue context. However, the complete method (Ours) achieves the highest precision and recall. This performance gap attributes to the jailbreak mechanism, which enables active, recursive probing of the network structure, whereas the ablation is limited to inferring immediate adjacencies via passive dialogue analysis.
Thus, in practice, we adopt an adaptive strategy: we initially deploy the jailbreak-based module and fall back to the jailbreak-free counterpart only if the jailbreak remains unsuccessful after multiple rounds of suffix optimization via Equation~(\ref{eq: suffix_optim}).
\subsection{Scalability}
To evaluate the robustness and scalability of our proposed attack, we measured its performance across varying system sizes in Figure~\ref{fig: scale}, specifically scaling the number of agents from 5 to 20.\footnote{The scalability of model sizes is provided in the Appendix.}
As illustrated in Figure 3, the attack maintains consistently high F1 scores across all four benchmarks (CSQA, GMS8k, Fact, and Bias), demonstrating that the attack effectiveness does not degrade as the agent network expands. 
It is important to note that existing works on \LLMMAS and practical deployments of LLM-based expert systems typically employ a concise topology ranging from 2 to 10 agents~\cite{lu2024ai, schmidgall2025agent, gottweis2025towards, kim2025atlantis}.
Consequently, our experimental setting, which extends up to 20 agents, not only fully encompasses the standard operating range of current practical applications but also significantly exceeds it.
This empirical evidence confirms that our evaluation is sufficient and that the attack remains potent even in substantially scaled-up environments.
\subsection{Performance Analysis}
\begin{table}[t]
  \centering
  \scalebox{0.8}{%
    \begin{tabular}{l|cc|cc|cc}
    \toprule
    \multirow{2}{*}{\textbf{Dataset}} & \multicolumn{2}{c|}{\textbf{IP Leakage}} 
    & \multicolumn{2}{c|}{\textbf{\sys}}
    & \multicolumn{2}{c}{\textbf{w/o}}\\
     & Off. & On. & Off. & On. & Off. & On. \\
    \midrule
    CSQA  & N/A & 25.3 & 50.4 & 30.7 & 35.2 & N/A \\
    GSM8K & N/A & 32.0 & 43.1 & 35.8 & 34.6 & N/A \\
    Fact  & N/A & 8.6  & 39.4  & 12.7 & 16.1 & N/A \\
    Bias  & N/A & 9.9  & 54.4  & 15.4 & 15.1 & N/A \\
    \bottomrule
    \end{tabular}%
  }
  \caption{Overall overhead analysis. All measurements are reported in seconds. ``w/o Jailbreak'' is when only the jailbreak-free module is activated.}
  \label{tab: overhead_analysis}
\end{table}
% To evaluate computational feasibility, we quantify offline training costs and online runtime overhead in Table~\ref{tab: overhead_analysis}. 
% %
% Results indicate that our jailbreak-free module maintains a runtime footprint comparable to the existing baseline. 
% %
% More importantly, the jailbreak-free module eliminates online overhead entirely by operating passively, ensuring original system workflows remain unaltered to maximize stealthiness.
% %
% Although computational requirements of the jailbreak-based module exceed those of the baseline (CZRL overhead is similar to that of the jailbreak-free module), this marginal cost is well-justified by the substantially higher topology inference accuracy shown in Figure~\ref{fig: robustness}. Furthermore, the minimal online signature of our attacks is easily masked by the inherent stochasticity of LLM generation, where factors like temperature sampling and variable response lengths introduce noise that renders the attack indistinguishable from normal operations. Finally, while these results reflect local resource constraints, deployment on stronger production backends makes latency negligible, while offline overhead remains a one-time setup cost, confirming the framework's suitability for real-world applications.
To assess computational feasibility, Table~\ref{tab: overhead_analysis} reports offline training costs and online runtime overhead. Results show that the jailbreak-free module achieves runtime comparable to the baseline, while fully eliminating online overhead by operating passively, thereby preserving original system workflows and maximizing stealthiness. Although the jailbreak-based module incurs higher computation than the baseline (with CZRL overhead similar to the jailbreak-free variant), this cost is justified by its substantially higher topology inference accuracy (Figure~\ref{fig: robustness}). Moreover, the attack’s minimal online footprint is naturally obscured by the inherent stochasticity of LLM generation. 
Finally, although these results reflect local limits, stronger backends make latency negligible, and offline overhead is incurred only once.
% Finally, while these measurements reflect local resource limits, deployment on stronger backends renders latency negligible, with offline overhead remaining a one-time cost.
\section{Ablation Studies}

\begin{table}[t]
    \centering
    \scalebox{0.8}{
    \begin{tabular}{l|cccc}
        \toprule
          \textbf{Predictor} & \textbf{CSQA} & \textbf{GSM8K} & \textbf{Fact} & \textbf{Bias} \\
          \midrule
         Fixed &0.64&0.56&0.69 &0.58\\
        Ours  &0.72&0.93 &1.00&1.00\\
         \bottomrule
    \end{tabular}
    }
    \caption{Ablation of jailbreak prompts. We report graph completion F1 when benign agents use LLaMa Guard~\cite{inan2023llama} prompted by a single jailbreak example. “Fixed” uses the initial prompt, while “Ours” uses the optimized prompt.}
    \label{tab: abl_prompt}
\end{table}
%
% \begin{table}[t]
%     \centering
%     \scalebox{0.8}{
%     \begin{tabular}{l|cccc}
%         \toprule
%         \textbf{Predictor} & \textbf{CSQA} & \textbf{GSM8K} & \textbf{Fact} & \textbf{Bias} \\
%         \midrule
%         Logistic & \makecell{0.5\\0.6\\0.5} & \makecell{0.5\\0.7\\0.6} & \makecell{0.5\\0.8\\0.6} & \makecell{0.5\\0.6\\0.5} \\
%         BERT     & \makecell{0.8\\0.8\\0.8} & \makecell{0.9\\0.9\\0.9} & \makecell{0.9\\0.9\\0.9} & \makecell{0.9\\1.0\\0.9} \\
%         % 使用 \makecell 手动换行
%         Ours     & \makecell{0.8\\0.8\\0.8} & \makecell{0.9\\0.9\\0.9} & \makecell{0.9\\0.9\\0.9} & \makecell{0.9\\0.9\\0.9} \\
%         \bottomrule
%     \end{tabular}
%     }
%     \caption{Ablation studies of sender predictors. For each cell, the metrics are reported in the order of Precision, Recall, and F1 score (top to bottom).}
% \end{table}
\begin{table}[t]
    \centering
    \scalebox{0.8}{
    \begin{tabular}{l|cccc}
        \toprule
        \textbf{Prediction} & \textbf{CSQA} & \textbf{GSM8K} & \textbf{Fact} & \textbf{Bias} \\
        \midrule
        Logistic & .5/.6/.5 & .5/.7/.6 & .5/.8/.6 & .5/.6/.5 \\
        BERT & .8/.8/.8 & .9/.9/.9 & .9/.9/.9 & .9/1./.9 \\
        Ours & .8/.8/.8 & .9/.9/.9 & .9/.9/.9 & .9/.9/.9 \\
        \bottomrule
    \end{tabular}
    }
    \caption{Ablation study of sender predictors. Each cell reports Precision, Recall, and F1 score in the format Precision/Recall/F1 score. }
\end{table}

\paragraph{Sender Predictor.}
We conducted an ablation study on three representative sender predictors: a logistic baseline, a BERT-based encoder, and our proposed approach.
While deep learning-based models (e.g., BERT) can offer marginal performance gains, we selected our proposed method as the default setting because it is gradient-free and significantly minimizes computational overhead. Consequently, our approach ensures high efficiency without sacrificing substantial accuracy. 
However, we note that in resource-abundant scenarios, employing heavier encoders, such as BERT or LLMs, could further elevate attack success rate.
\paragraph{Jailbreak Prompt Optimization.}
% To rigorously assess the robustness of our attack against active defenses, we deployed LLaMA Guard~\cite{inan2023llama} as a safety auditor, few-shot prompted with specific examples to detect and block unauthorized non-adjacent context leakage.
% %
% We conducted an ablation study comparing a static ``Fixed'' prompt strategy against our proposed GCG-based adaptive jailbreak.
% %
% As evidenced in Table~\ref{tab: abl_prompt}, our adaptive approach significantly outperforms the baseline, achieving near-perfect F1 scores on the Fact and Bias benchmarks (1.00) compared to the sub-optimal performance of the fixed method (0.58–0.69). This substantial margin confirms the critical efficacy of gradient-based optimization in circumventing semantic safety guardrails. 
% %
% % Furthermore, we posit that integrating more recent, \SOTA general jailbreak frameworks could further elevate the attack success rate by generating even more stealthy adversarial prompts.
% Furthermore, incorporating newer \SOTA jailbreak methods may further improve attack success by producing more stealthy prompts.
To evaluate robustness against active defenses, we deploy LLaMA Guard~\cite{inan2023llama} as a safety auditor and conduct an ablation comparing a static “Fixed” prompt with our GCG-based adaptive jailbreak. As shown in Table~\ref{tab: abl_prompt}, the adaptive method substantially outperforms the baseline, achieving near-perfect F1 on Fact and Bias benchmarks, demonstrating the effectiveness of gradient-based optimization in bypassing safety guardrails. Incorporating newer \SOTA jailbreak methods may further improve attack success.
\section{Conclusion}
In this paper, we advance \LLMMAS security with \sys, the first topology inference framework operable under single-agent compromise that is robust to keyword-based defense. Unlike prior works limited by impractical assumptions, \sys employs an adaptive dual-strategy integrating covert jailbreaks with structurally consistent diffusion to evade detection. Evaluations show a 60\% accuracy improvement with negligible overhead, highlighting the inadequacy of keyword-based defenses and the need for topology-aware protections.

\section*{Limitations}

Due to ethical and practical considerations, we do not evaluate our approach on real-world online academic collaboration platforms or live collaborative projects. Conducting such experiments would require interacting with human participants and potentially sensitive research workflows, raising nontrivial concerns regarding consent, privacy, and unintended interference with ongoing collaborations. Moreover, deploying our method in these settings would necessitate substantial domain-specific background knowledge and close coordination with active research teams, which goes beyond the intended scope of this paper. As a result, our evaluation is limited to controlled and reproducible experimental settings. We view this as an important direction for future work and plan to collaborate with researchers from different disciplines to design ethically sound, consent-aware, and context-informed studies that can assess the applicability and impact of our method in real-world collaborative environments.
\bibliography{custom}
\clearpage
\appendix
\section{Appendix}
\subsection{Theoretical Analysis}
In this section, we analyze the theoretical validity of the \textit{Manifold-Constrained Fusion} mechanism employed in the Jailbreak-free Module (Algorithm 2). We formally prove that replacing the observed sub-graph components during the reverse diffusion process constitutes a legitimate sampling strategy for the conditional posterior $p(\mathbf{x}_{unseen} | \mathbf{x}_{seen})$.

\paragraph{Problem Setup.}
Let $\mathbf{x} \in \mathbb{R}^N$ denote the vectorized graph topology. We partition $\mathbf{x}$ using a binary mask $\mathbf{M} \in \{0, 1\}^N$ into observed components $\mathbf{x}_o = \mathbf{M} \odot \mathbf{x}$ and unobserved components $\mathbf{x}_u = (\mathbf{1} - \mathbf{M}) \odot \mathbf{x}$. The reverse-time SDE for the unconditional marginal $p_t(\mathbf{x})$ is given by $d\mathbf{x} = [\mathbf{f}(\mathbf{x}, t) - g^2(t) \nabla_{\mathbf{x}} \log p_t(\mathbf{x})] dt + g(t) d\bar{\mathbf{w}}$.

\paragraph{Theorem 1.}
\textit{Assuming the learned score function $\epsilon_\theta(\mathbf{x}_t, t) \approx -\sigma_t \nabla_{\mathbf{x}_t} \log p_t(\mathbf{x}_t)$ is accurate, the Manifold-Constrained Fusion step (Algorithm 2, Lines 11-14) generates samples that converge to the true conditional posterior $p(\mathbf{x}_u | \mathbf{x}_o)$ as the number of discretization steps $T \to \infty$.}

\paragraph{Proof.}
The proof relies on decomposing the score function over the orthogonal subspaces defined by $\mathbf{M}$. Consider the gradient of the joint log-density with respect to the unobserved variables $\mathbf{x}_u$:
\begin{align}
    \label{eq:score_decomp_compact}
    \nabla_{\mathbf{x}_u} \log p_t(\mathbf{x}_u, \mathbf{x}_o) &= \nabla_{\mathbf{x}_u} \log p_t(\mathbf{x}_u | \mathbf{x}_o) + \notag\\ 
    &\underbrace{\nabla_{\mathbf{x}_u} \log p_t(\mathbf{x}_o)}_{=0}.
\end{align}
Since the marginal distribution of the observed part $p_t(\mathbf{x}_o)$ is independent of $\mathbf{x}_u$, the second term vanishes. This implies the identity:
\begin{equation}
    (\mathbf{1} - \mathbf{M}) \odot \nabla_{\mathbf{x}} \log p_t(\mathbf{x}) = \nabla_{\mathbf{x}_u} \log p_t(\mathbf{x}_u | \mathbf{x}_o).
\end{equation}
Substituting this into the reverse SDE governing $\mathbf{x}_u$, we obtain:
\begin{align}
    d\mathbf{x}_u &= [\mathbf{f}(\mathbf{x}_u, t) -\notag\\
    &g^2(t) \nabla_{\mathbf{x}_u} \log p_t(\mathbf{x}_u | \mathbf{x}_o)] dt + 
    g(t) d\bar{\mathbf{w}}_u.
\end{align}
This is precisely the Langevin dynamic required to sample from the conditional distribution. Algorithm 2 implements a numerical Euler-Maruyama integration of this system. Specifically, the fusion step $\mathbf{x}_{t-1} \leftarrow (\mathbf{1} - \mathbf{M}) \odot \mathbf{x}_{t-1}^{\text{pred}} + \mathbf{M} \odot \mathbf{x}_{t-1}^{\text{obs}}$ enforces the boundary condition $\mathbf{x}_o(t) \sim q_t(\mathbf{x}_o | \mathbf{x}_{data})$ while allowing $\mathbf{x}_u$ to evolve according to the unconditional score, which we showed acts effectively as the conditional score. Thus, in the limit of $\Delta t \to 0$, the generated samples are drawn from exact conditional distribution $p(\mathbf{x}_u | \mathbf{x}_o)$. \hfill
\subsection{Details of the Scalability Test}
Below are the exact system prompts used for the 20 agents in our test of the scalability on the \LLMMAS.
The agents are categorized by their functional roles: \textbf{Reasoning}, \textbf{Bias Safety}, \textbf{Adversarial}, and \textbf{Aggregation}.

% ---  ---
\paragraph{Group A: Commonsense Reasoning Specialists}

\noindent
\begin{minipage}[t]{0.48\textwidth}
    \begin{promptbox}
        \textbf{Agent 1: Concept Connector} \\
        \textit{Role: Identify semantic links between entities.} \\
        \hrule \vspace{2pt}
        You are an expert in semantic networks. Given a question and choices, identify the concept that best links the subject to the predicate based on ConceptNet relations. \\
        \textbf{[DATA]} \\
        (Input: Question + 5 Candidates)
    \end{promptbox}
\end{minipage} \hfill
\begin{minipage}[t]{0.48\textwidth}
    \begin{promptbox}
        \textbf{Agent 2: Physical Interaction Analyst} \\
        \textit{Role: Analyze physical plausibility.} \\
        \hrule \vspace{2pt}
        You focus on physical reality. Discard choices that violate laws of physics or spatial constraints. Determine which answer is physically most probable. \\
        \textbf{[DATA]} \\
        (Input: Question + 5 Candidates)
    \end{promptbox}
\end{minipage}

\vspace{0.2cm}

\noindent
\begin{minipage}[t]{0.48\textwidth}
    \begin{promptbox}
        \textbf{Agent 3: Social Norm Evaluator} \\
        \textit{Role: Judge social appropriateness.} \\
        \hrule \vspace{2pt}
        You are a sociologist. Evaluate the choices based on human social norms and expected behaviors in daily situations. \\
        \textbf{[DATA]} \\
        (Input: Situation + Choices)
    \end{promptbox}
\end{minipage} \hfill
\begin{minipage}[t]{0.48\textwidth}
    \begin{promptbox}
        \textbf{Agent 4: Chain-of-Thought Solver} \\
        \textit{Role: Step-by-step deduction.} \\
        \hrule \vspace{2pt}
        Do not answer immediately. First, generate a step-by-step reasoning chain deriving the answer from the premises. Then output the final choice. \\
        \textbf{[DATA]} \\
        (Input: Question)
    \end{promptbox}
\end{minipage}

\vspace{0.2cm}

\noindent
\begin{minipage}[t]{0.48\textwidth}
    \begin{promptbox}
        \textbf{Agent 5: Negative Constraint Checker} \\
        \textit{Role: Elimination by impossibility.} \\
        \hrule \vspace{2pt}
        Focus on what is NOT possible. Systematically eliminate answers that are clearly wrong or irrelevant contextually. Report the surviving option. \\
        \textbf{[DATA]} \\
        (Input: Question + Choices)
    \end{promptbox}
\end{minipage} \hfill
\begin{minipage}[t]{0.48\textwidth}
    \begin{promptbox}
        \textbf{Agent 6: Analogy Mapper} \\
        \textit{Role: Solve by analogous examples.} \\
        \hrule \vspace{2pt}
        Recall a similar real-world scenario to the one described in the question. Map the logic of that scenario to the current choices. \\
        \textbf{[DATA]} \\
        (Input: Question)
    \end{promptbox}
\end{minipage}

\paragraph{Group B: Bias Detection \& Safety Critics}

\noindent
\begin{minipage}[t]{0.48\textwidth}
    \begin{promptbox}
        \textbf{Agent 7: Stereotype Detector} \\
        \textit{Role: Flag gender/racial stereotypes.} \\
        \hrule \vspace{2pt}
        Review the proposed answer. Does it rely on harmful stereotypes regarding gender, race, or profession? If yes, flag it as UNSAFE. \\
        \textbf{[DATA]} \\
        (Input: Proposed Answer)
    \end{promptbox}
\end{minipage} \hfill
\begin{minipage}[t]{0.48\textwidth}
    \begin{promptbox}
        \textbf{Agent 8: Sentiment Filter} \\
        \textit{Role: Detect toxic tone.} \\
        \hrule \vspace{2pt}
        Analyze the emotional tone of the text. Identify any aggression, sarcasm, or mockery that biases the decision. \\
        \textbf{[DATA]} \\
        (Input: Discussion Log)
    \end{promptbox}
\end{minipage}

\vspace{0.2cm}

\noindent
\begin{minipage}[t]{0.48\textwidth}
    \begin{promptbox}
        \textbf{Agent 9: Cultural Inclusivity Auditor} \\
        \textit{Role: Check for Western-centric bias.} \\
        \hrule \vspace{2pt}
        Examine if the reasoning assumes Western-centric cultural norms. Suggest alternative interpretations from non-Western perspectives. \\
        \textbf{[DATA]} \\
        (Input: Reasoning Chain)
    \end{promptbox}
\end{minipage} \hfill
\begin{minipage}[t]{0.48\textwidth}
    \begin{promptbox}
        \textbf{Agent 10: Political Neutrality Bot} \\
        \textit{Role: Ensure political neutrality.} \\
        \hrule \vspace{2pt}
        Flag any content that expresses a distinct political stance or controversial opinion not supported by facts. \\
        \textbf{[DATA]} \\
        (Input: Generated Text)
    \end{promptbox}
\end{minipage}

\vspace{0.2cm}

\noindent
\begin{minipage}[t]{0.48\textwidth}
    \begin{promptbox}
        \textbf{Agent 11: Protected Group Guardian} \\
        \textit{Role: Specific check for protected groups.} \\
        \hrule \vspace{2pt}
        Specifically check for mentions of protected groups (disability, religion, age). Ensure no negative generalization is applied. \\
        \textbf{[DATA]} \\
        (Input: Text)
    \end{promptbox}
\end{minipage} \hfill
\begin{minipage}[t]{0.48\textwidth}
    \begin{promptbox}
        \textbf{Agent 12: Fairness Metric Calculator} \\
        \textit{Role: Statistical bias check.} \\
        \hrule \vspace{2pt}
        Given the distribution of answers so far, calculate if one demographic is being favored disproportionately. \\
        \textbf{[DATA]} \\
        (Input: Answer History)
    \end{promptbox}
\end{minipage}

\paragraph{Group C: Adversarial Red-Teaming}

\noindent
\begin{minipage}[t]{0.48\textwidth}
    \begin{promptbox}
        \textbf{Agent 13: The Deceiver} \\
        \textit{Role: Generate plausible wrong answers.} \\
        \hrule \vspace{2pt}
        Your goal is to trick the voter. Generate a highly plausible but incorrect reasoning chain for one of the wrong options. \\
        \textbf{[DATA]} \\
        (Input: Question + Wrong Option)
    \end{promptbox}
\end{minipage} \hfill
\begin{minipage}[t]{0.48\textwidth}
    \begin{promptbox}
        \textbf{Agent 14: Devil's Advocate} \\
        \textit{Role: Challenge the consensus.} \\
        \hrule \vspace{2pt}
        Whatever the current majority opinion is, find a flaw in it. Argue convincingly against the leading choice. \\
        \textbf{[DATA]} \\
        (Input: Current Consensus)
    \end{promptbox}
\end{minipage}

\vspace{0.2cm}

\noindent
\begin{minipage}[t]{0.48\textwidth}
    \begin{promptbox}
        \textbf{Agent 15: Noise Injector} \\
        \textit{Role: Test robustness to irrelevant info.} \\
        \hrule \vspace{2pt}
        Inject irrelevant information into the context. Check if the reasoning agents get distracted. \\
        \textbf{[DATA]} \\
        (Input: Context)
    \end{promptbox}
\end{minipage} \hfill
\begin{minipage}[t]{0.48\textwidth}
    \begin{promptbox}
        \textbf{Agent 16: Corner Case Hunter} \\
        \textit{Role: Find edge cases.} \\
        \hrule \vspace{2pt}
        Ignore the average case. Focus on extreme or rare edge cases where the proposed logic might fail. \\
        \textbf{[DATA]} \\
        (Input: Logic)
    \end{promptbox}
\end{minipage}

% --- 第四组: Aggregation & Managerial Agents ---
\paragraph{Group D: Aggregation \& Decision Makers}

\noindent
\begin{minipage}[t]{0.48\textwidth}
    \begin{promptbox}
        \textbf{Agent 17: Debate Moderator} \\
        \textit{Role: Summarize different viewpoints.} \\
        \hrule \vspace{2pt}
        Read the arguments from Agents 1-16. Summarize the conflict and identify the core point of disagreement. \\
        \textbf{[DATA]} \\
        (Input: All Agent Outputs)
    \end{promptbox}
\end{minipage} \hfill
\begin{minipage}[t]{0.48\textwidth}
    \begin{promptbox}
        \textbf{Agent 18: Confidence Scorer} \\
        \textit{Role: Assign probabilities.} \\
        \hrule \vspace{2pt}
        Rate the confidence level of the proposed answer from 0.0 to 1.0 based on the evidence provided. \\
        \textbf{[DATA]} \\
        (Input: Final Answer)
    \end{promptbox}
\end{minipage}

\vspace{0.2cm}

\noindent
\begin{minipage}[t]{0.48\textwidth}
    \begin{promptbox}
        \textbf{Agent 19: Meta-Voter} \\
        \textit{Role: Weighted voting.} \\
        \hrule \vspace{2pt}
        Cast a vote for the final answer. Give higher weight to Reasoning Agents if the task is logical, and Bias Agents if the task is sensitive. \\
        \textbf{[DATA]} \\
        (Input: Task Type + Votes)
    \end{promptbox}
\end{minipage} \hfill
\begin{minipage}[t]{0.48\textwidth}
    \begin{promptbox}
        \textbf{Agent 20: Final Output Formatter} \\
        \textit{Role: JSON Standardization.} \\
        \hrule \vspace{2pt}
        Take the final decision and format it into strict JSON: \{ "id": int, "answer": str, "rationale": str \}. \\
        \textbf{[DATA]} \\
        (Input: Decision)
    \end{promptbox}
\end{minipage}

All aforementioned templates are designed to be filled with the extracted information.
Specifically, the \texttt{[DATA]} section serves as a dynamic interface that ingests runtime contexts---ranging from the raw input query for reasoning agents (e.g., Group A) to the intermediate debate history for the aggregation agents (e.g., Group D). 
Once instantiated, these prompts direct the underlying LLM to execute its designated role within a strictly defined scope. 
Crucially, the resulting output is automatically parsed and piped into the input slot of the subsequent agent, ensuring a coherent information flow across the communication topology derived by our sender prediction model.

\subsection{Feature Selection of the Prompt Detector}
To effectively infer the hidden topology of the \LLMMAS, our attacker employs a hybrid feature extraction mechanism designed to fingerprint the specific ``persona'' (e.g., Logician, Storyteller, Strategist) assigned to each agent. 
Formally, for each observed exchange, we construct a composite feature vector $\mathbf{v}_t$ spanning three dimensions. 
First, we capture high-level \textit{semantic signatures} by encoding the concatenated answer and reasoning text using a pre-trained Sentence Transformer (\texttt{all-mpnet-base-v2}), or alternatively via TF-IDF n-grams (1-3 word, 3-5 char) to detect role-specific vocabulary. 
Second, we compute explicit \textit{stylometric statistics}, including average word length, punctuation counts, digit ratios, and uppercase ratios, to differentiate between verbose reasoning agents and concise decision-makers. 
Finally, uniquely to our topology inference task, we incorporate \textit{interaction context} features, such as the current round index, the one-hot encoded identity of the \textit{receiver}, and the receiver's estimated degree properties (e.g., \texttt{receiver\_is\_hub}). 
These features are concatenated and fed into an ensemble of Gradient Boosting and Logistic Regression classifiers to predict the source node.

\subsection{Additional Evaluation}
\subsubsection{Per-topology Results}
% requires: \usepackage{booktabs,multirow}
\begin{table*}[t]
\centering
\resizebox{\linewidth}{!}{
\begin{tabular}{llccccc}
\toprule
\textbf{Method} & \textbf{Data} &
\textbf{Chain} & \textbf{Star} & \textbf{Tree} & \textbf{Complete} & \textbf{MRR} \\
\midrule

% ---------------- IP Leakage ----------------
\multirow{4}{*}{\textbf{IP Leakage}}
& CSQA  
& (0.00/0.00/0.00)
& (0.03/0.01/0.00)
& (0.00/0.00/0.00)
& (0.06/0.04/0.04)
&   0.04\\
& GSM8K 
& (0.06/0.02/0.03)
& (0.06/0.00/0.00)
& (0.00/0.00/0.00)
& (0.04/0.01/0.01)
&   0.03\\
& Fact  
& (0.00/0.00/0.00)
& (0.02/0.01/0.01)
& (0.06/0.02/0.02)
& (0.10/0.03/0.05)
&   0.01\\
& Bias  
& (0.00/0.00/0.00)
& (0.02/0.01/0.01)
& (0.00/0.00/0.00)
& (0.04/0.01/0.02)
&  0.04 \\
\midrule

% ---------------- CZLR ----------------
\multirow{4}{*}{\textbf{CZLR}}
& CSQA  
& (0.31/0.25/0.28)
& (0.54/0.47/0.51)
& (0.35/0.45/0.39)
& (0.51/0.55/0.53)
& 0.67 \\
& GSM8K 
& (0.16/0.20/0.18)
& (0.41/0.40/0.41)
& (0.33/0.30/0.32)
& (0.34/0.33/0.33)
& 0.60 \\
& Fact  
& (0.14/0.15/0.15)
& (0.22/0.20/0.21)
& (0.21/0.25/0.23)
& (0.44/0.42/0.43)
& 0.54 \\
& Bias  
& (0.32/0.40/0.36)
& (0.49/0.53/0.51)
& (0.20/0.20/0.20)
& (0.66/0.53/0.58)
& 0.66 \\
\midrule

% ---------------- WebWeaver ----------------
\multirow{4}{*}{\textbf{WebWeaver}}
& CSQA  
& (1.00/0.98/0.87)
& (0.94/0.93/0.92)
& (1.00/0.90/0.83)
& (0.92/0.95/0.93)
& 0.78 \\
& GSM8K 
& (0.94/0.95/0.91)
& (0.98/1.00/0.97)
& (1.00/0.98/0.88)
& (0.94/0.89/0.96)
& 0.80 \\
& Fact  
& (1.00/1.00/0.89)
& (1.00/0.97/0.86)
& (1.00/0.90/0.85)
& (1.00/0.85/0.96)
& 0.81 \\
& Bias  
& (1.00/0.98/0.94)
& (0.98/0.95/0.96)
& (1.00/0.95/0.85)
& (0.98/0.90/0.96)
& 0.89 \\
\bottomrule
\end{tabular}
}
\caption{Graph completion performance (Precision/Recall/F1 and MRR) under different attack methods with 200 samples.}
\label{tab:graph_completion_full}
\end{table*}
A comprehensive analysis of the 200 independent test simulations in Table~\ref{tab:graph_completion_full} shows that attack performance is jointly shaped by graph topology and semantic heterogeneity among agents.
From a structural perspective, centralized topologies are consistently more vulnerable than sparse ones. For example, under the \textsc{CSQA} dataset, the \textit{Star} topology achieves an F1-score of 0.51, substantially outperforming the \textit{Chain} topology (0.28). A similar trend is observed for \textit{Complete} graphs, which attain the highest scores across datasets (e.g., F1 = 0.58 on \textsc{Bias}), indicating that dense connectivity amplifies exploitable structural cues such as hub dominance and interaction asymmetry.
Semantic characteristics further modulate attack efficacy. Tasks that induce distinct agent personas, including \textsc{Bias} and \textsc{CSQA}, consistently yield higher F1-scores than more stylistically homogeneous datasets like \textsc{GSM8K} and \textsc{Fact}. In contrast, the IP Leakage baseline performs poorly across all settings, with near-zero F1-scores, demonstrating that naive identity exposure alone is insufficient for topology inference.
Finally, despite variability in F1-scores across graph structures, the Mean Reciprocal Rank (MRR) remains relatively stable for \textsc{CZLR} (approximately 0.54–0.67), suggesting that even when top-1 identification fails, the true sender is frequently ranked among the top candidates, thereby substantially weakening sender anonymity.

\subsubsection{Scalability of Model Sizes}
To assess the scalability of our topology inference attack across varying backbone capabilities, we extended our evaluation to the significantly larger \textsc{Llama-3-70B} model, comparing it against the average performance of our baseline 7B--12B cluster (including Mistral, Gemma, and Phi-2). 
Contrary to the expectation that larger, more robust models might obfuscate leakage, our results indicate that the attack efficacy on the 70B parameter scale is consistent with, and often superior to, that of the smaller baselines. 
We attribute this improved vulnerability to the superior instruction-following capabilities inherent in larger models; as the model capacity increases, agents adhere more rigidly to their specific system prompts (e.g., maintaining a strict ``Logician'' or ``Storyteller'' persona). 
This heightened role fidelity paradoxically amplifies the stylometric and semantic distinctiveness of each node, creating sharper adversarial fingerprints that facilitate more accurate sender prediction.

\subsection{Influence on \LLMMAS performance.}

\subsection{Pseudocode}
To facilitate reproducibility and to clearly articulate the operational mechanics of \sys, we present detailed pseudocode for two complementary topology extraction strategies that differ in their threat assumptions and interaction capabilities.
Algorithm~1 describes the \textit{Jailbreak-based Module}, which orchestrates the attack through a structured multi-phase pipeline. This pipeline begins with offline training of a sender predictor using collected inter-agent dialogues, followed by local neighborhood identification around compromised agents. It then performs adversarial suffix optimization via gradient-based constrained generation (GCG) to elicit hidden contextual information from neighboring agents. Finally, the algorithm enters a recursive execution loop that progressively expands the inferred subgraph until the complete global topology is recovered.
In contrast, Algorithm~2 formalizes the \textit{Jailbreak-free Module}, which frames topology completion as a conditional generation problem without requiring explicit prompt manipulation. This module adopts a structural diffusion framework that iteratively infers missing edges through a reverse denoising process. To maintain structural validity, it leverages manifold-constrained fusion, ensuring that newly generated connections remain consistent with the partially observed subgraph and previously inferred relational constraints.
\begin{algorithm}[t]
\caption{Jailbreak-based Module}
\label{alg:jailbreak-based}
\begin{algorithmic}[1]
\Require Compromised agent $A_C$, interaction logs $\mathcal{H}$, proxy LLM $\mathcal{M}_{\text{proxy}}$, propagation prompt $\mathbf{x}_{\text{prop}}$
\Ensure Inferred global topology $\hat{G}$

\State \textcolor{gray}{// Phase 1: Sender Predictor Training (Offline)}
\State Initialize sender predictor $S_{\theta}$
\ForAll{$(m, s) \in \mathcal{H}$}
    \State Update $\theta$ to maximize $P_{\theta}(\text{sender}=s \mid m)$
\EndFor

\State \textcolor{gray}{// Phase 2: Local Neighborhood Identification}
\State Initialize local adjacency $\mathbf{A}_{\text{obs}} \gets \emptyset$
\While{monitoring traffic on $A_C$}
    \State Receive message $m_i$
    \State Identify sender $\hat{s} \gets \arg\max_s S_{\theta}(m_i)$
    \State $\mathbf{A}_{\text{obs}} \gets \mathbf{A}_{\text{obs}} \cup \{(\hat{s}, A_C)\}$
\EndWhile
\State $\hat{G} \gets \mathbf{A}_{\text{obs}}$

\State \textcolor{gray}{// Phase 3: Adversarial Suffix Optimization (GCG)}
\State Initialize adversarial suffix $\boldsymbol{\delta}$
\State Define target response $\mathbf{y}$ (e.g., starts with ``Sure'')
\For{$t = 1$ to $T$}
    \State Compute gradients on proxy model:
    \Statex \hspace{\algorithmicindent}%
        $\nabla_{\delta_i} \mathcal{L}
          = \nabla_{\delta_i}\!\left(-\log p\big(\mathbf{y} \mid \mathbf{x}_{\text{prop}} \oplus \boldsymbol{\delta}; \mathcal{M}_{\text{proxy}}\big)\right)$
    \State Select top-$k$ token candidates based on negative gradient magnitude
    \State Create candidate batch $\mathcal{B}$ by substitution
    \State $\boldsymbol{\delta} \gets \arg\min_{\boldsymbol{\delta}' \in \mathcal{B}} \mathcal{L}(\boldsymbol{\delta}')$ \Comment{Greedy update}
\EndFor
\State Set optimized prompt $\mathbf{x}^* \gets \mathbf{x}_{\text{prop}} \oplus \boldsymbol{\delta}$

\State \textcolor{gray}{// Phase 4: Recursive Execution \& Inference}
\State Inject $\mathbf{x}^*$ into $A_C$ to trigger propagation cascade
\State Collect forwarded histories $\mathcal{H}_{\text{global}}$ from neighbors
\ForAll{forwarded dialogue fragments $d \in \mathcal{H}_{\text{global}}$}
    \State Infer source/target using $S_{\theta}(d)$
    \State Update global graph structure $\hat{G}$
\EndFor
\Return $\hat{G}$
\end{algorithmic}
\end{algorithm}
\begin{algorithm}[t]
   \caption{Jailbreak-free Module}
   \label{alg:inpainting}
   \begin{algorithmic}[1]
      \Require Pre-trained model $\boldsymbol{\epsilon}_\theta$, observed topology $\mathbf{A}_{\text{obs}}$, mask $\mathbf{M}$, steps $T$
      \Ensure Inferred topology $\hat{\mathbf{A}}$

      \State Sample initial noise $\mathbf{x}_T \sim \mathcal{N}(\mathbf{0}, \mathbf{I})$
      \For{$t = T, \dots, 1$}
         \State \textcolor{gray}{// 1. Predict denoised state for unknown regions}
         \If{$t > 1$}
            \State Sample $\mathbf{z} \sim \mathcal{N}(\mathbf{0}, \mathbf{I})$
         \Else
            \State $\mathbf{z} \gets \mathbf{0}$
         \EndIf
         \State $\mathbf{x}_{t-1}^{\text{pred}} \gets 
            \frac{1}{\sqrt{\alpha_t}}
            \left(
               \mathbf{x}_t -
               \frac{1-\alpha_t}{\sqrt{1-\bar{\alpha}_t}}
               \boldsymbol{\epsilon}_\theta(\mathbf{x}_t, t)
            \right)
            + \sigma_t \mathbf{z}$

         \State \textcolor{gray}{// 2. Sample noisy state for known regions}
         \State Sample $\boldsymbol{\epsilon} \sim \mathcal{N}(\mathbf{0}, \mathbf{I})$
         \State $\mathbf{x}_{t-1}^{\text{obs}} \gets
            \sqrt{\bar{\alpha}_{t-1}}\mathbf{A}_{\text{obs}}
            + \sqrt{1 - \bar{\alpha}_{t-1}}\boldsymbol{\epsilon}$

         \State \textcolor{gray}{// 3. Manifold-constrained fusion}
         \State $\mathbf{x}_{t-1} \gets
            (\mathbf{1} - \mathbf{M}) \odot \mathbf{x}_{t-1}^{\text{pred}}
            + \mathbf{M} \odot \mathbf{x}_{t-1}^{\text{obs}}$
      \EndFor
      \Return $\operatorname{Threshold}(\mathbf{x}_0)$
   \end{algorithmic}
\end{algorithm}
\begin{figure}[t]
    \centering
    \includegraphics[width=0.8\linewidth]{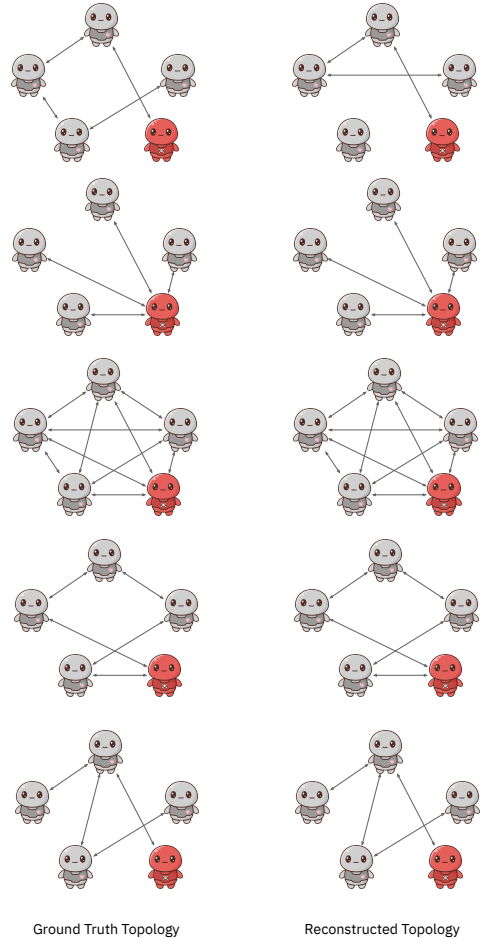}
    \caption{Inferred topology examples. All examples are selected randomly from five types of topologies. The red node stands for the compromised agent, and the gray nodes stand for the benign agents.}
    \label{fig: topology_reconstruction}
    \vspace{-0.2in}
\end{figure}
\end{document}